\documentclass[fleqn,usenatbib]{mnras}

\usepackage{newtxtext,newtxmath}

\usepackage[T1]{fontenc}
\usepackage{ae,aecompl}

\usepackage{mathtools}
\usepackage{graphicx}	
\usepackage{float}
\usepackage{amsmath}	

\usepackage{subcaption}

\graphicspath{{./}{figures/}}

\usepackage{xspace}

\newcommand{\MS}{\ensuremath{\mathrm{M}_\odot}\xspace}

\newcommand{\lcdm}{{$\Lambda$}CDM}               

\usepackage[dvipsnames]{xcolor}
\usepackage[normalem]{ulem} 

\title[]{Reversing Arrested Development: A New Method to Address Halo Assembly Bias}

\author[Smith et al.]{
William J. Smith$^{1}$\thanks{E-mail: william.j.smith@vanderbilt.edu (WJS)},
Andreas A. Berlind$^{2}$,
Manodeep Sinha$^{3, 4}$
\\
$^{1}$Department of Physics and Astronomy, Vanderbilt University, Nashville, TN 37235, USA \\
$^{2}$Division of Astronomical Sciences, National Science Foundation, Alexandria, VA22314, USA\\
$^{3}$Centre for Astrophysics \& Supercomputing, Swinburne University of Technology, Melbourne, Australia\\
$^{4}$ARC Centre of Excellence for All Sky Astrophysics in 3 Dimensions (ASTRO 3D), Australia}

\date{Accepted XXX. Received YYY; in original form ZZZ}

\pubyear{2023}

\begin{document}
\label{firstpage}
\pagerange{\pageref{firstpage}--\pageref{lastpage}}
\maketitle

\begin{abstract}
Halo assembly bias is a phenomenon whereby the clustering of dark matter halos is dependent on halo properties, such as age, at fixed mass. Understanding the origin of assembly bias is important for interpreting the clustering of galaxies and constraining cosmological models. One proposed explanation for the origin of assembly bias is the truncation of mass accretion in low-mass halos in the presence of more massive halos, called ‘arrested development’. Halos undergoing arrested development would have older measured ages and exhibit stronger clustering than equal mass halos that have not undergone arrested development. We propose a new method to test the validity of this explanation for assembly bias, and correct for it in cosmological N-body simulations. The method is based on the idea that the early mass accretion history of a halo, before arrested development takes effect, can be used to predict the late-time evolution of the halo in the absence of arrested development. We implement this idea by fitting a model to the early portion of halo accretion histories and extrapolating to late times. We then calculate "corrected" masses and ages for halos based on this extrapolation and investigate how this impacts the assembly bias signal. We find that correcting for arrested development this way leads to a factor of two reduction in the strength of the assembly bias signal across a range of low halo masses. This result provides new evidence that arrested development is a cause of assembly bias and validates our approach to mitigating the effect.
\end{abstract}

\begin{keywords}
cosmology: theory - dark matter - galaxies: formation - galaxies: haloes - large-scale structure of universe - methods: numerical
\end{keywords}



\section{Introduction} \label{Introduction}
Halo assembly bias is a phenomenon observed in simulations of large-scale structure in which the clustering of dark matter halos depends on the properties of their assembly histories in addition to their masses \citep{Croton_2007,Wechsler_Tinker2018}. Assembly bias has been found using many secondary halo properties, notably age \citep{Sheth_Tormen_2004,Gao_2005,Wechsler_2006,Wang_2007,Li_Mo_Gao_2008}, concentration \citep{Wechsler_2006,Gao_White_2007,Faltenbacher_2010,Lazeyras_2017,Villarreal_2017} and spin \citep{Gao_White_2007,Faltenbacher_2010,Lacerna_2012,Lazeyras_2017,Villarreal_2017,Johnson_2019}, among others. Halo assembly bias is important to understand because it challenges a basic assumption that most statistical models of the galaxy-halo connection are founded upon - the assumption that the clustering of dark matter halos depends {\em only} on halo mass. Thus, understanding assembly bias has important implications for accurately modeling galaxy clustering in simulations, modeling galaxy clustering empirically, explaining observed galaxy clustering, and constraining cosmological parameters~\citep{Wechsler_Tinker2018}.

After assembly bias was discovered for multiple secondary properties in multiple simulations, many works have attempted to explain its physical origins and uncover the subsequent implications for hierarchical clustering models and models of the galaxy-halo connection. It is currently believed that multiple concurrent factors contribute to assembly bias. In their article investigating causes of assembly bias, \citet{Mansfield2020} pinpoint four concurrent sources of assembly bias. One cause, random Gaussian field statistics, impacts the high halo mass regime \citep{Dalal2008}. Three causes impact the low mass regime: splashback radius definition \citep{Sunayama2016,Diemer_2021}, gravitational heating, and halo arrested development \citep{Dalal2008,Hahn2009,Salcedo2018}. 

Though \citet{Mansfield2020} present three causes of low-mass assembly bias in low-mass halos, each manifests from the same underlying cause - arrested development. Arrested development is the truncation or slow-down of mass growth in some lower mass halos due to tidal forces induced by the presence of a more massive neighbor (or a denser overall halo environment) (\citealt{Dalal2008}, \citealt {Hahn2009}, \citealt{Salcedo2018}). Splashback halos (or flyby halos) pass within the inner regions of more massive halos, which causes the truncation of mass accretion, and even mass stripping, due to this encounter (e.g., \citealt{Sinha_2012}, \citealt{Wetzel_2014}). Gravitational heating occurs when a low-mass halo cannot accrete particles in its vicinity because they are too energetic. Those particles are more energetic because a more massive neighboring halo accelerates them. Thus, a method that addresses arrested development as a truncation of mass accretion due to a nearby more massive neighbor or neighbors addresses all three of these proposed causes of low-mass assemble bias simultaneously.

This work investigates arrested development as the primary cause of halo assembly bias.  \citet{Hahn2009} first tested the hypothesis of arrested development as a cause of assembly bias using an N-body simulation and analyzing the correlation between halo assembly history and environment, in this case defined not by a density field in the form of a two-point correlation function, but instead by a tidal/shear field. They argued that these tidal effects do indeed suppress halo growth. Though they did not go so far as to conclude that arrested development is the cause of assembly bias, they did conclude that it is an important factor contributing to its emergence in simulation data.

Another work lending evidence to arrested development as a source of halo assembly bias is \citet{Hearin2016}. They analyzed halo mass accretion rates of pairs of halos as functions of the individual pair masses, pair separation, and pair environment. They found mass accretion conformity out to many times the halos' virial radii. Halo pairs in overall denser regions have less mass accretion, and the lower the halos' masses, the more pronounced this conformity. They state that this conformity finding caused by the halo tidal environment is an alternative quantification of other assembly bias findings. 

A more recent and more direct investigation of sources of assembly bias is presented by \citet{Salcedo2018}, who analyzed a multitude of halo properties to disentangle which properties have the strongest effects on assembly bias. One of their main analyses studied halo properties as a function of distance to a halo's nearest more massive neighbor. Importantly, they found that, if one controls for `neighbor bias', then bins halos by mass and splits by age or concentration, the assembly bias signal is mostly removed. However, they used several different halo properties as both primary and secondary properties to bin and split halos and found that controlling for neighbor bias does not successfully remove assembly bias in all cases, complicating their ability to discern a simple cause. 

Though multiple explanations for assembly bias have been proposed, none alone has satisfactorily explained the phenomenon. \citet{Mansfield2020} attempted to assign a relative importance and a cohesive model for all three of the low-mass causes of assembly bias. They used a single set of simulations with multiple prescriptions and proxies for calculating splashback radii.  They measured tidal strength from both the largest nearby halo and the general strength of the aggregate tidal field, and measure tidal heating. From this, they determined the relative contributions of each factor by finding the percentage of halos that need to be removed for a given proxy to eliminate the assembly bias signal. The results of \citet{Mansfield2020} support the connection between halo tidal forces halo assembly bias in agreement with the conclusions of 
\citet{Hahn2009}, \citet{Hearin2016}, and \citet{Salcedo2018}, but they express a caveat that assembly bias is likely not an effect that can be approximated from the tidal force of a halo's single most massive neighbor alone.

We propose a new approach to investigate arrested development as a cause of assembly bias. The idea is to identify arrested development in halos and predict the growth they would have had if they had not undergone a slow-down or truncation of mass accretion. Since arrested development mostly occurs at late times, when massive neighbors have collapsed and produce strong tidal fields, we can attempt to extrapolate from the early part of a halo's history to predict its late evolution. To perform this extrapolation, we adopt prior work in using a functional form to fit mass accretion histories. Having removed the effects of arrested development from halo mass accretion histories, we can investigate to what extent the assembly bias signal is reduced.

This new approach offers more than an investigation of the cause of halo assemble bias; it also may offer a straightforward way of removing the halo assembly bias signal in simulations consistent with its underlying physical cause. Though prior work has been well motivated in addressing physical causes of assembly bias, the low-mass causes of assembly bias propose challenges for strategies to mitigate this bias. Because our proposed method attempts to correct the underlying cause of assembly bias - arrested development - it could be a useful tool that can be applied to any simulation, with practical applications, such as producing assembly bias-free mock catalogs.

Section~\ref{Data} provides an overview of the simulation and how we extract mass accretion histories. Section~\ref{Fitting} describes our procedure for fitting mass accretion histories and correcting for arrested development. Section~\ref{clustering_section} shows how we calculate the halo assembly bias signal via their two-point correlation function. Section~\ref{Results} shows our main results from attempting to mitigate arrested development and investigating how this impacts the halo assembly bias signal. Section \ref{summary} discusses limitations and implications for future assembly bias investigations.

\section{Data} \label{Data}

\subsection{The Vishnu Simulation}
Vishnu is a large-scale cosmological high-resolution N-body simulation. The simulation contains $1000$ snapshots of particles from $z = 99$ to $z = 0$ and was evolved using the GADGET-2 N-body TreeSPH algorithm \citep{Springel_Gadget2005} in a \lcdm{} cosmology with $h = 0.70$, $\Omega_{m} = 0.25$, $\Omega_{\Lambda} = 0.75$ and $\sigma_{8} = 0.8$. The initial power spectrum of Vishnu was calculated using CAMB, a Python code for cosmological calculations \citep{Lewis_2000}. Initial positions and velocities were determined using the 2LPT code \citep{Scoccimarro_1998} at $z = 99$. Vishnu has a boxsize of 130 $h^{-1}$Mpc and a particle mass of $3.2151 \times 10^{7} h^{-1} \MS$ \citep{Johnson_2019}. Halos were identified using the ROCKSTAR algorithm \citep{Behroozi_rockstar} with the virial overdensity set to $\Delta_{vir}$ \citep{Bryan_Norman_1998}. Halo mergertrees were constructed with the Consistent-Trees algorithm of \citet{Behroozi_mergertree}. 

\begin{figure*}
\centering
\includegraphics[scale=0.85]{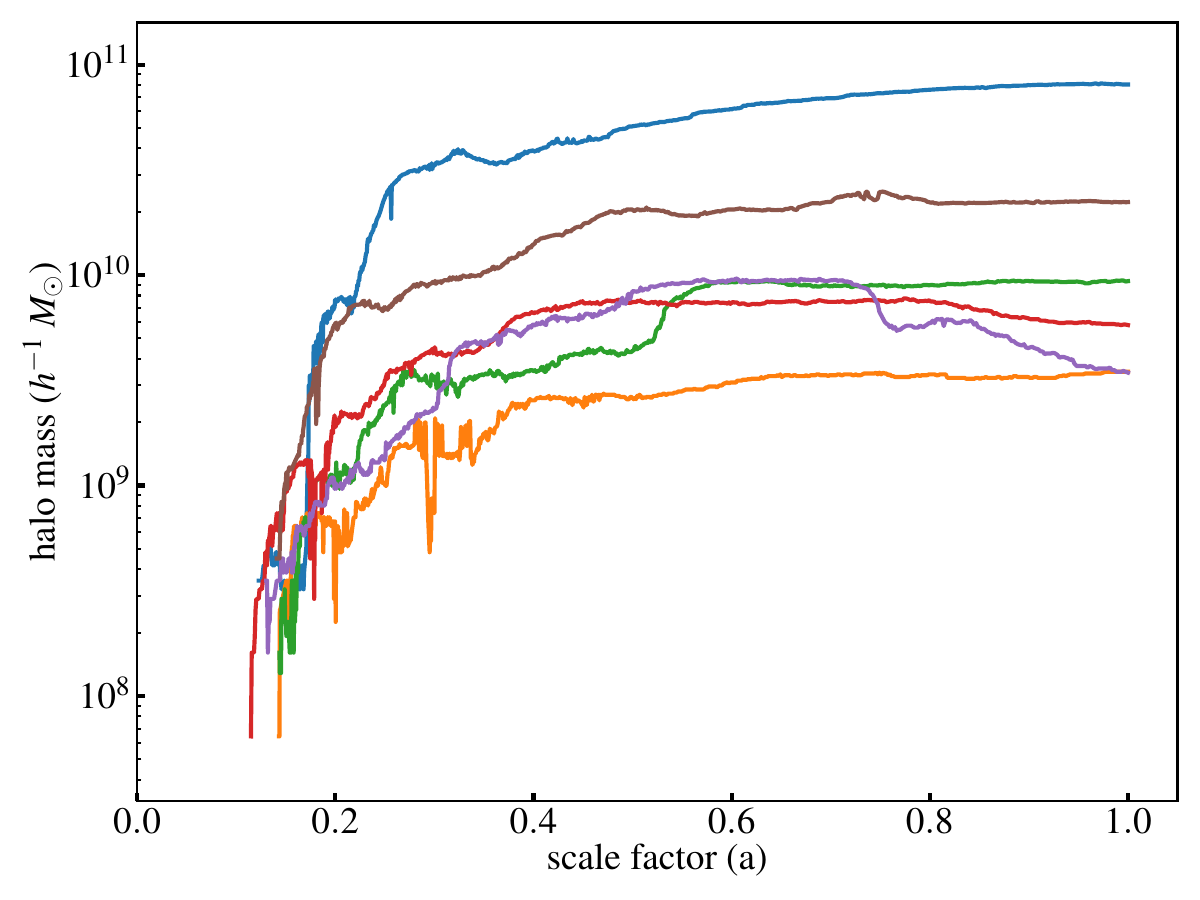}
\caption{Six mass accretion histories from the Vishnu simulation. Each line shows mass as a function of scale factor, $a$, for a particular randomly chosen halo. The halos exhibit a large variety in their histories. For example, the halo represented in purple shows a mass loss in its later history ($a\sim0.75$). Most show a leveling off of mass accretion in their late histories.}
\label{fig:sample_mass_accretion_histories}
\end{figure*}

The Vishnu simulation offers a high mass and temporal resolution compared to most other cosmological N-body simulations while still maintaining a large enough volume to measure the large-scale clustering of halos. This makes it an ideal simulation for studying the connection between halo mass accretion histories and assembly bias. Since this analysis only uses the total mass of halos and their clustering, and is not concerned with internal properties like density profiles, we are confident in using halos with a fairly small number of particles. However, since the focus of this work is to analyze mass accretion histories, not just halos in the present, we require halos to be resolved at very early times when they were much less massive. This means that our final $z=0$ mass threshold should be much higher than the minimum mass we can resolve. We adopt a final halo mass threshold for our analysis of $10^{10} h^{-1} \MS$, which corresponds to about 300 particles. This allows us to resolve halo histories over the period where they have grown by at least an order of magnitude. If, for example, a present-day halo of mass $10^{10} \ h^{-1} \ \MS{}$ has a half-mass age at scale factor $a=0.5$, then the halo at this earlier point would have roughly $150$ particles, still enough for the analysis in this work and enough that we can probe its accretion history back to even earlier times.

The main idea in this paper is to fit the early portion of the mass accretion history (MAH) of halos, before arrested development has taken effect, and extrapolate the fit to predict how the halos would have grown in the absence of arrested development. We thus need to adopt a truncation point in time from which to extrapolate. This point should be at a time that is early enough that arrested development has not yet taken place, but late enough that the halo is well resolved in a sufficient number of previous timesteps to give us confidence in the early accretion history fit. The earliest truncation point that we adopt in this analysis is a=0.5, which satisfies these requirements. We expect that halos that experience arrested development will have extrapolated final masses that are higher than their actual final masses in the simulation. Therefore, in order to have a complete sample of halos with extrapolated final masses greater than $10^{10} h^{-1} \MS$, we must perform fits on halos with actual final masses that are lower than that so that arrested development halos are included in the sample. To satisfy this requirement, we create a sample of all halos in the simulation with final $z=0$ masses greater than $10^{9.5} \ h^{-1} \ \MS{}$, which provides enough of a buffer to ensure a complete sample above $10^{10} h^{-1} \MS$. Additionally, we adopt a high mass cutoff of $10^{11.6} \ h^{-1} \ \MS{}$ because, above this threshold, there are too few halos per mass bin to calculate accurate clustering statistics. Finally, only halos classified as central halos at $z = 0$ are used in this analysis (i.e., not subhalos). These conditions result in a sample of roughly $1.2 \times 10^{6} $ halos from the Vishnu simulation that we fit for the analysis.

\begin{figure}
    \centering
    \includegraphics[scale=0.5]{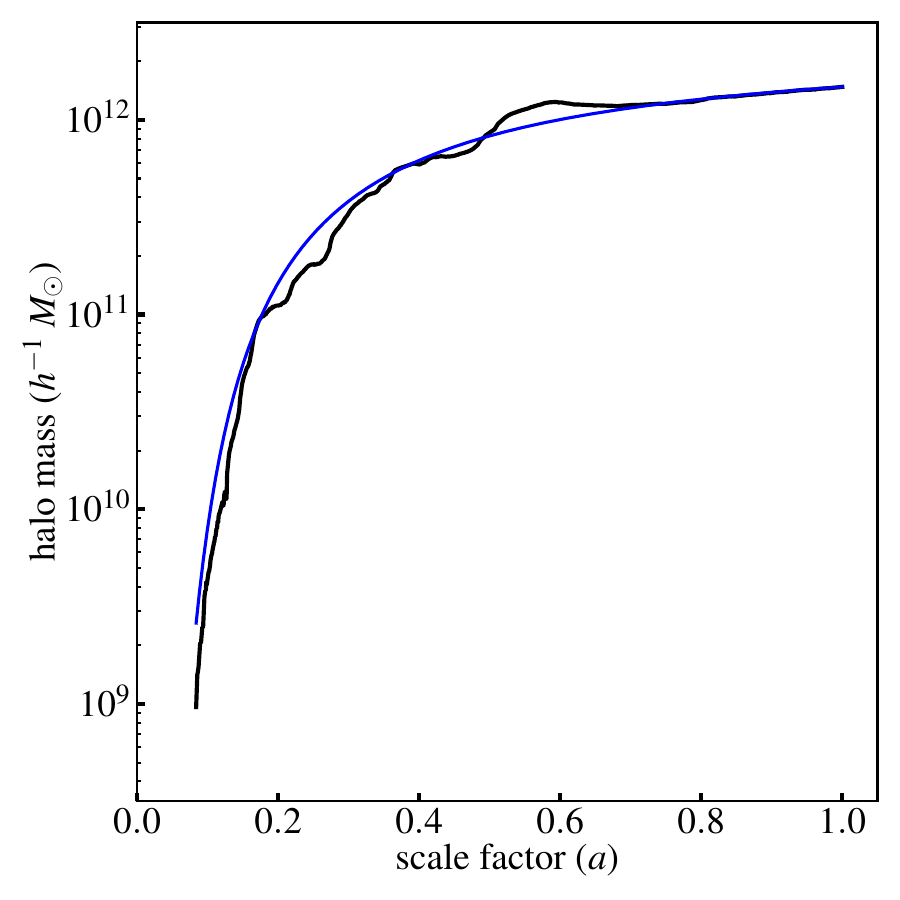}
    \caption{Example of a mass accretion history, always following the most massive progenitor back in time, for a halo from the Vishnu simulation (black), fit with an exponential function of the form $M(a) = M_{0} e^{(-\alpha z)}$ (solid blue). This exponential fit is a simple parameterization that fits many halos well.}
    \label{fig:new_Wechsler}
\end{figure}

\subsection{Extracting the Mass Accretion Histories}

The Consistent-Trees algorithm of \citet{Behroozi_mergertree} creates halo mergertrees in Vishnu, which include every branch of every progenitor halo that results in a current halo. Since every halo at every timestep in the simulation is given a unique ID number, the Consistent-Trees algorithm records the descendent halo IDs for every halo. From this, the mass accretion histories of every halo at $a = 1$ are extracted by moving backwards in time and finding progenitor halo IDs and selecting the progenitor with the most mass. One branch is randomly chosen in the rare event that multiple progenitor halos have the same mass. This process is repeated until a halo with no progenitor halo is reached. We refer to the scale factor at which the halo with no progenitor is reached as the ``epoch of appearance" of that halo. We note that this does not mean that the halo did not exist at earlier times, but that it had too few particles to be registered by the ROCKSTAR halo finder algorithm. Figure~\ref{fig:sample_mass_accretion_histories} shows six randomly selected mass accretion histories from Vishnu. The halos exhibit a large variety in their detailed histories. For example, the halo represented by the purple line exhibits sudden mass loss at around $a = 0.75$. However, all of these example halos share roughly the same overall form. In their early histories, the halos accrete at a rapid fractional rate, while later in their histories, the mass accretion histories level out.

\begin{figure*}

\begin{subfigure}{\textwidth}
    \centering
    \includegraphics[width=0.8\textwidth]{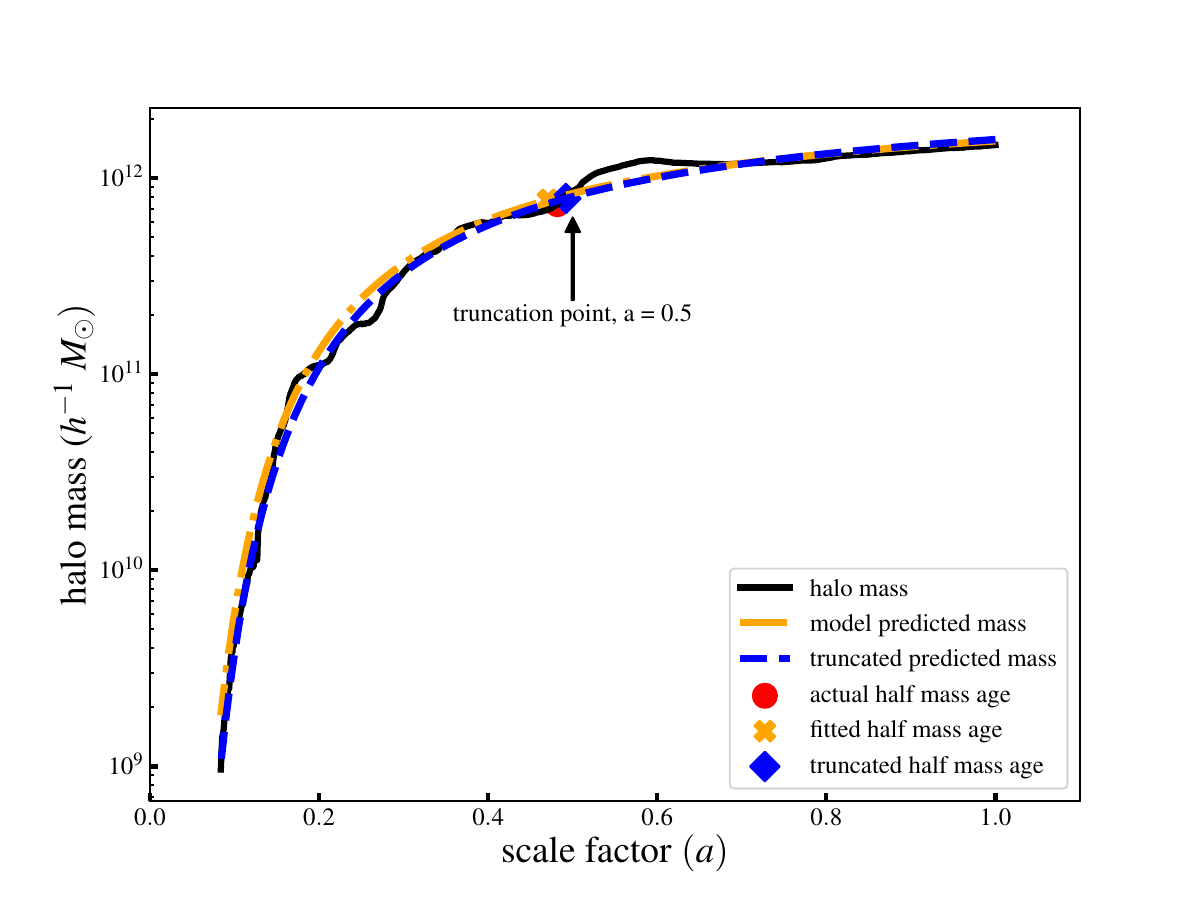}
    \label{fig:reg_fit_real_halo}
\end{subfigure}

\begin{subfigure}{\textwidth}
    \centering
    \includegraphics[width=0.8\textwidth]{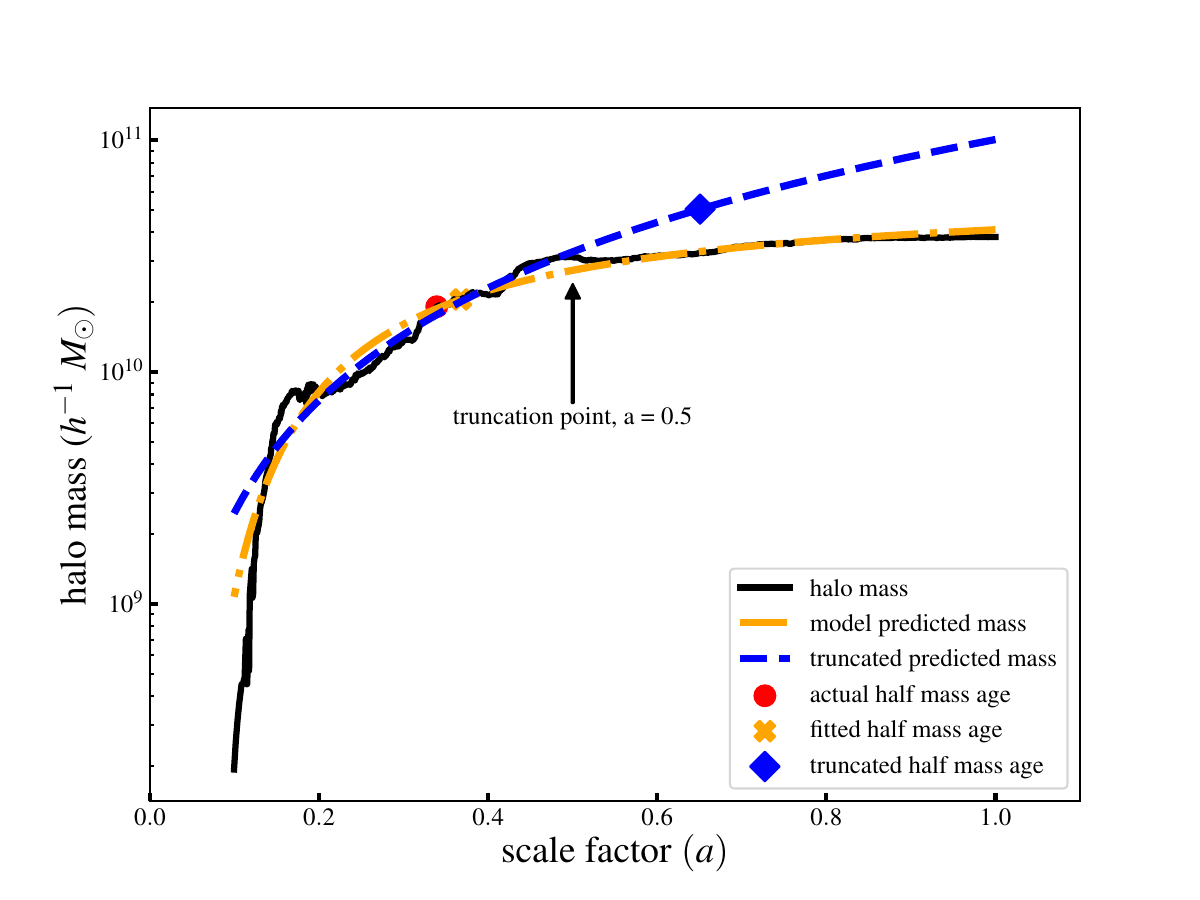}
    \label{fig:AD_fit_real_halo}
\end{subfigure}

\caption{Two fits for a randomly chosen `regular' halo (top panel) and arrested development halo (bottom panel) in the Vishnu simulation. The solid black curves are the actual mass accretion histories of the halos as extracted from the simulation. The orange (dot-dashed) curves show the best-fit model accretion histories, using the fitting function in equation~\ref{McBride_two_param}, when the fit is performed over the entire mass accretion histories. The blue (dashed) curves show the fit results when the fit is only performed over the earlier portion of the histories, up to scale factor $a = 0.5$. This scale factor is denoted by the black arrows. The blue curves at later times thus represent an extrapolation from the earlier history fits. The large points show the scale factors corresponding to the half-mass ages of the three histories. Fitting these halos through both methods results in nearly identical fits with similar final predicted masses and ages for the 'regular' halo (top panel), but very different fits for the arrested development halo (bottom panel). In this latter case, the fitting function accurately predicts the final halo mass and age when fit over the entire history, but it predicts a higher final halo mass and a younger age when fit over only the early history.} 
\label{fig:regular_and_AD_halo_fits}
\end{figure*}

\section{Fitting the Mass Accretion Histories} \label{Fitting}

A search for a generalized mass accretion history function that can be derived from an analytic foundation and verified by simulations is ongoing. It has been recognized that, in general, most halos fractionally accrete much of their mass early, then experience a period of gradually slower fractional accretion (\citealt{Ludlow_2013}) (see Figure  \ref{fig:sample_mass_accretion_histories}), with this effect being more or less pronounced depending on other halo properties (notably overall halo mass) \citep{Wechsler2002,Tasitsiomi_2004,McBride_2009}. Early work attempted to derive a generalized halo mass accretion history profile from mass assembly histories (e.g. \citealt{Avila_Reese_1998}, \citealt{Ryden_Gunn_1987}). In a seminal paper, \citet{Wechsler2002} extracted and analyzed structural merger trees of halos from simulation data. Prior work by \citet{Bullock_2001} had used a simulation to study the density profiles of dark matter halos, and \citet{Wechsler2002} attempted to find a correlation between these density profiles and the assembly history of the halos. In doing so, they proposed an exponential fitting function for mass accretion histories of the form:
\begin{equation}\label{one_param}
M(a) = M_{0} \mathrm{e} ^ {-\alpha z},
\end{equation}
where $M(a)$ is the mass of the halo at some scale factor $a$, $M_{0}$ is the final mass of the halo, $z$ is the redshift which is related to the scale factor as $a = (1 + z)^{-1}$, and $\alpha$ is a free parameter. Figure~\ref{fig:new_Wechsler} shows this functional form, fit to one of our Vishnu halos. \citet{Wechsler2002} notably restricted fitting this function to halos above $10^{12} \ h^{-1} \ M_{\odot}$ at $z = 0$ to have reliable fits for most halos. This choice created a better set of halo fits for their analysis, but it raises doubt on this function’s appropriateness for fitting halos of lower masses.

Though this functional fit for mass accretion histories is simple and powerful, subsequent researchers recognized its limitations and suggested improved functional forms. For example, \citet{VDB_2002} suggested a two-parameter function and \citet{Tasitsiomi_2004} proposed a two-parameter, more general function of: 
\begin{equation}
\label{tasitsiomi_two_param}
\Tilde{M}(\Tilde{a}) = \Tilde{a}^{p} \ \exp\left(\alpha\left(1 - \frac{1}{\Tilde{a}}\right)\right),
\end{equation}
in which $\Tilde{M}(\Tilde{a}) = \frac{M}{M_{0}}$, $\Tilde{a} = \frac{a}{a_{0}}$, and $M_{0}$ and $a_{0}$ are the virial mass and the scale factor of the halo at the time (or epoch) of observation, and $p$ and $\alpha$ are free parameters. This parameterization allows halos to be fit with a functional form that is a combination of an exponential of $\frac{1}{\Tilde{a}}$  and a power law of base $\Tilde{a}$. For cases in which $p=0$, this function for fitting mass accretion histories simplifies to the one-parameter exponential function of \citet{Wechsler2002}. \citet{Tasitsiomi_2004} proposed and applied this form to 14 halos for analysis. \citet{McBride_2009} then tested this form over a much larger sample of approximately $500,000$ halos from the Millennium Simulation and found it to be "a reasonable fit". They rewrote the function identically as:
\begin{equation}
    \label{McBride_two_param}
    M(z) = M_{0}(1 + z)^{\alpha} \mathrm{e}^{\beta z}
\end{equation}
with $\alpha$ and $\beta$ as free parameters, which is the form used for this analysis, though we convert redshift $z$ to scale factor $a$ for this work. Some sources using this two-parameter functional form have a negative exponential term, though this is just a sign convention choice. Furthermore, \citet{McBride_2009} suggest a four-tiered classification system based on the fitted values of the two free parameters to classify halo types. This classification scheme refers to halos with $\alpha < 0.35$ as `good exponential' halos, halos with $\alpha - \beta < -0.45$ as `steep late growth' halos, halos with $-0.45 < \alpha - \beta < 0$ as `shallow late growth' halos, and halos with $\alpha - \beta > 0$ as `late plateau/decline' halos. Though this categorization is not used explicitly in this analysis, it is a consideration for future work.

This specific two-parameter functional fit was further investigated by \citet{Wong_Taylor_2012}, who noted a concern that for most fits $\alpha$ and $\beta$ are correlated and exhibit some degeneracy (meaning multiple value combinations of $\alpha$ and $\beta$ can produce the same functional fit). They further proposed a model where the primary free parameter is $(\beta-\alpha)$. \citet{Correa2015} explained this correlation and provided a more comprehensive justification for this functional form by fitting to an analytic mass history model based on the EPS formalism and connecting $\alpha$ and $\beta$ to the initial power spectrum. Though continued work investigates the best functional fit for mass accretion histories, this is currently the most well-researched and tested function for this purpose.

We use the parameterization of \citet{McBride_2009} (equation~\ref{McBride_two_param}) to functionally model all the mass accretion histories in our simulation: once over the entire history and again over a portion of the early history out to specified truncation points. The fit over the entire history checks how well the function models the mass accretion over the entire history of the halos and serves as a comparison to the functional fit over the early histories of the halos. The equation being fit is equation \ref{McBride_two_param}, where $M_0$, $\alpha$, and $\beta$ are free parameters. Since the independent time variable is in units of scale factor instead of redshift, the function is converted to be a function of scale factor $a$ instead of redshift $z$. Though this function is identical to the functional form of mass accretion histories used by \citet{Correa2015} and prior works, one difference is that we add a third free parameter, the final halo mass. In past works, the final mass $M_0$ was assumed to be known and fixed to the final mass of the halo in the simulation data. However, in this analysis we assume that the final halo mass, had the halo not undergone arrested development, is not known from the simulation itself but can be inferred from extrapolating the early halo accretion data. Thus, this is the first investigation into halo accretion that allows $M_{0}$ to be a free parameter when fitting halos. 

\begin{figure*}
\centering
\begin{subfigure}{0.45\textwidth}
    \includegraphics[width=\textwidth]{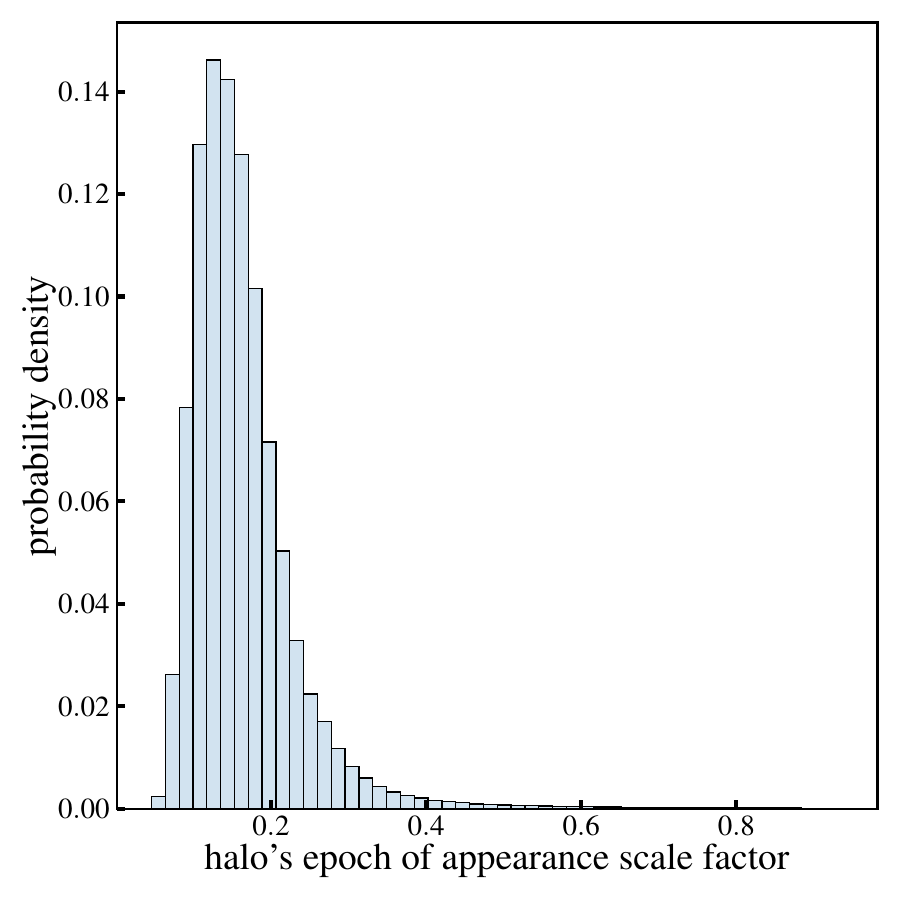}
    \label{fig:Vishnu_halo_appearance_times}
\end{subfigure}
\hfill
\begin{subfigure}{0.45\textwidth}
    \includegraphics[width=\textwidth]{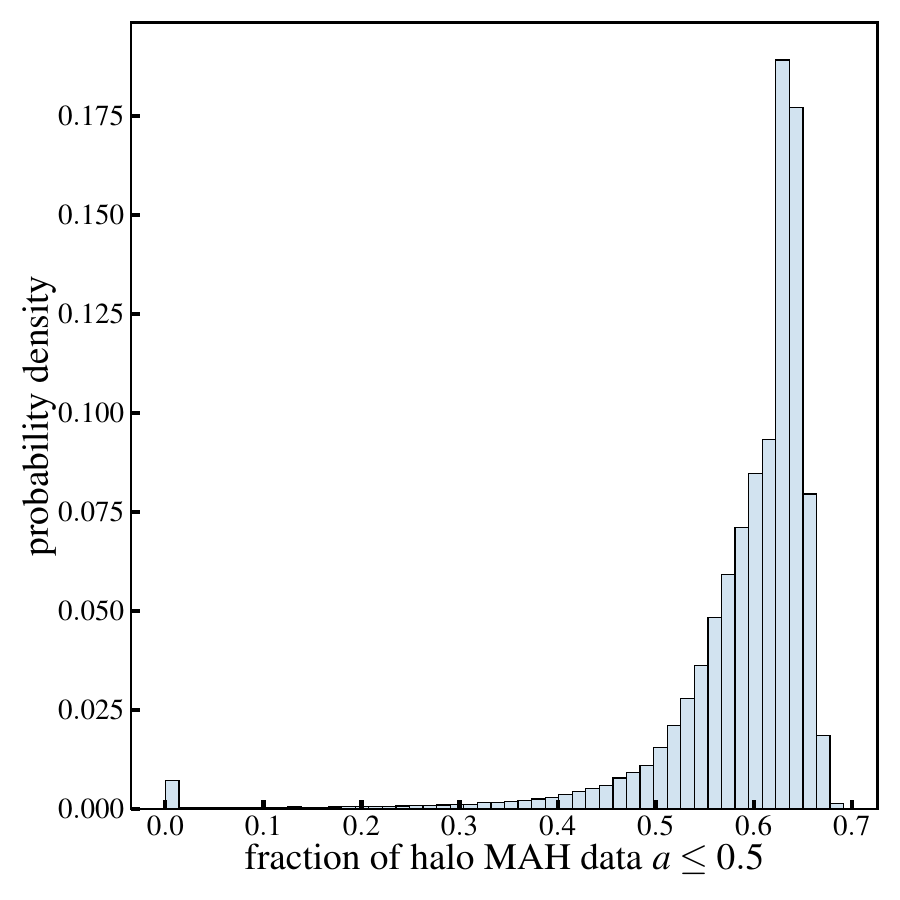}
    \label{fig:MAHS_proportion_before_05}
\end{subfigure}
\caption{Tests to check whether halos have enough of their history in place before the projected scale factor of $a=0.5$, in order to robustly fit this early history and extrapolate halo growth to late times. {\it Left panel:} the distribution of halo "epochs of appearance", which is the earliest scale factor when the progenitor of a present-day halo was first resolved in the Vishnu simulation (following the main, most massive, trunk of the merger tree back in time). The distribution shows that the vast majority of halos first appear well before $a=0.5$, meaning that we are not excluding a significant number of halos from our analysis due to them not existing before our chosen projected scale factor. {\it Right panel:} the distribution of the fraction of time steps in Vishnu halo histories that occur before $a=0.5$. The distribution shows that, for almost all halos, a large fraction of their resolved history occurs before our chosen projected scale factor (with a peak near 60\%). This means that our fits to the early histories of halos are based on a large enough number of epochs to be robust. The results in both panels apply to all halos above $10^{9.5} \ h^{-1} \ M_{\odot}$.}
\label{fig:Halo_fitting_checks}
\end{figure*}


The fit is performed using the \textit{optimize} sub-package of the \textit{SciPy} package. For each halo fit, boundary conditions are imposed on the free parameters. Specifically, the parameter $M_{0}$, being a mass, is constrained to be positive, and both $\alpha$ and $\beta$ are constrained to only be negative. In the case of $\beta$, this constraint arises from the functional form. A positive $\beta$ would cause the function to have an asymptote at positive infinity as it approaches scale factor zero regardless of the values of the other two parameters, representing an unphysical halo with infinite mass at the beginning of the universe. The $\alpha$ parameter has no physical motivation for imposing a boundary condition; however, a negative value of $\alpha$ imposes a restriction against mass loss for the halo. Importantly, for some halo history forms, a range of positive values for $\alpha$ can also result in mass accretion histories that do not exhibit mass loss, but the specific value of $\alpha$ for which the halo mass function has a ‘turnover point’ and exhibits mass loss is itself dependent on the value of $\beta$. For this work, a boundary condition is imposed that $\alpha$ be negative, preventing all halo fits from exhibiting mass loss. We restrict the possibility of mass loss when fitting halos because it more accurately accounts for the arrested development hypothesis, which states that assembly bias is caused by halos not accreting mass as they would without tidal forces. Whether a halo undergoing arrested development gains mass more slowly than it otherwise would or whether it actually loses mass, in the absence of the tidal forces that result in arrested development, halos can only gain mass. Thus, we expect that, whether the halo itself loses mass or not, the "corrected" mass history from our functional fit should only increase in mass.

Once the optimal values of the three parameters of $M_{0}$, $\alpha$, and $\beta$ are calculated using Scipy’s curve fit module, a model halo's fitted mass history can be calculated from the function and parameters. From the model mass accretion history, the model half mass age is calculated as the scale factor of the model halo when it has acquired half its final mass. Figure~\ref{fig:regular_and_AD_halo_fits} shows the results of this fit for two example halos from the Vishnu simulation, one "normal" halo (top panel) and one arrested development halo (bottom panel). The solid black lines show the actual mass accretion histories of the halos as extracted from the simulation. The dot-dashed orange curves show the best-fit model accretion histories, using the fitting function in equation~\ref{McBride_two_param}, when the fit is performed over the entire mass accretion histories. The half-mass ages, estimated from the actual and fitted histories, are denoted by the red circle and orange X symbols, respectively. In both cases, the fitted history tracks the overall shape of the actual history fairly well and has a similar final mass and half-mass age.

Next, mass accretion histories are fit using only the early portion of their histories. This results in a best-fit analytic history with new best-fit values of $M_{0}$, $\alpha$, and $\beta$. The later part of the halo’s history can be extrapolated from the fit to the earlier portion of the history. In this work, we refer to the latest epoch that is used in the fit as the ‘projected scale factor.' We begin with a projected scale factor of $a = 0.5$. This value was chosen because it is early enough in most halo histories to allow the extrapolation to correct for arrested development but late enough that halos are resolved in a sufficient number of earlier epochs of the simulation in order for a trustworthy fit to be made. The blue curves in Figure~\ref{fig:regular_and_AD_halo_fits} show halo fits using a projected scale factor of 0.5 for the examples of "normal" and arrested development halos. The corresponding half-mass ages from these fits are denoted by the blue diamond symbols. The figure shows that in the case of the "normal" halo, this new fit matches the previous fit well, meaning that the late-time extrapolation of the mass accretion history represents the true history well even though it is only based on the early history of the halo. In the case of the arrested development halo, however, the extrapolated accretion history overpredicts the late-term mass of the halo by a large factor, with a $z=0$ mass that is more than two times higher than the actual mass of the halo. Likewise, the half-mass scale factor of the "corrected" history is $\sim$50\% larger (i.e., younger) than the original age. Our interpretation of these differences is that the early history of the halo is not affected by arrested development and so the extrapolation of the fit to this early history is predicting how the halo would have grown in the absence of arrested development. 

We can use Figure~\ref{fig:regular_and_AD_halo_fits} to demonstrate the importance of the chosen projected scale factor. As the projected scale factor (represented in Figure~\ref{fig:regular_and_AD_halo_fits} by the black arrows) is set to earlier times, the number of simulation time steps over which to fit the halo between its appearance in the simulation and its truncation point decreases. Conversely, as the projected scale factor moves to later time steps, the truncated halo fit (the blue curve) naturally moves closer to the model halo fit (the orange curve). In this case, the projected halo converges to the model halo fit, undermining the goal of accounting for arrested development in the halo. In addition to a projected scale factor of $a = 0.5$, we also analyze projected scale factors of $a = 0.6$, $a = 0.7$, $a = 0.8$, and $a = 0.9$. We fit each halo from its scale factor of first appearance in the simulation to the projected scale factor. Halos that formed after the projected scale factor are excluded from the analysis. 

Figure \ref{fig:Halo_fitting_checks} demonstrates the appropriateness of our chosen projected scale factor of $a = 0.5$. In choosing a projected scale factor, we strive to ensure two criteria. First, most halos in the simulation should have an epoch of appearance before the projected scale factor. Clearly, halos whose oldest resolved progenitor halos appear after our chosen projected scale factor cannot be fit. Such halos would be excluded by our analysis, potentially biasing the result. The left panel of Figure~\ref{fig:Halo_fitting_checks} shows the distribution of epoch of appearance for all halos with final masses above $10^{9.5}\ h^{-1} \ \MS{}$ in the Vishnu simulation. The overwhelming majority of halos appeared before $a = 0.5$, with most appearing just before $a = 0.2$. Thus, we are confident that our choice in the projected scale factor of $a = 0.5$ does not entail a significant source of bias due to excluded halos. Second, we look at the data in each mass accretion history, focusing on how many of the simulation time steps fall before the projected scale factor. If there are too few timesteps between the halo's epoch of appearance and the projected scale factor, we risk extrapolating the halo's late history from too brief a period in the halo's history, yielding unreliable results. Conversely, if most of the timesteps are between the halo's epoch of appearance and the projected scale factor, then there is little to extrapolate. In this case, the truncated halo will yield the same result as the model fit. The right panel of Figure~\ref{fig:Halo_fitting_checks} shows the distribution of the fraction of the timesteps between the epoch of appearance of the halo and the projected scale factor of $a = 0.5$. To help understand this, consider the sharp peak at roughly $0.6$. This signifies that for a large number of halos, about 60\% of the timesteps in the simulation fall between the halo's epoch of appearance and the projected scale factor of $a = 0.5$, while 40\%  of the timesteps fall between $a = 0.5$ and $a = 1$. For most halos, the truncated fits hit the optimal middle ground in which we are not extrapolating over too little or not extrapolating at all.

After halos are fitted, we return the final halo mass from the simulation, the model final mass (the best-fit parameter $M_{0}$ fitted over the entire halo), and a projected final mass (the best-fit parameter $M_{0}$ fitted over only the halo's early history). Only halos above $10^{10} \ h^{-1} \ \MS{}$ are analyzed for each of the three mass categories. However, we fit all halos down to final masses of $10^{9.5} \ h^{-1} \ \MS{}$ to allow for the halos that have actual masses below $10^{10} \ h^{-1} \ \MS{}$ but may have model and/or projected masses above $10^{10} \ h^{-1} \ \MS{}$ to be included in the analysis, as not including these halos would bias the model and projected halo mass data. The $10^{9.5} \ h^{-1} \ \MS{}$ mass cut threshold was chosen to provide a wide enough mass range to reasonably include most halos that fall into this category but without compromising the reliability of the halo fits from low-mass limits on halo resolution. We support this choice by noting that only roughly 5\% of the model halos with a final mass greater than $10^{10} \ h^{-1} \ M_{\odot}$ were fitted from halos with an original final mass between $10^{9.5}$ and $10^{10} \ h^{-1} \ M_{\odot}$, and only 0.3 \% of model halos with a mass greater than $10^{10} \ h^{-1} \ M_{\odot}$ had original halo masses less than $10^{9.8} \ h^{-1} \ M_{\odot}$. 

We define half-mass age separately for each of the three mass accretion history types. The half-mass age of the halo is the point at which, following the most massive progenitor along the halo's merger tree, the halo reaches half its present mass. The model half-mass age is the point at which the function fitted over the entire halo reaches half the final fitted mass, and the projected half-mass age is the point at which the projected function, fitted over the early portion of the mass accretion history, reaches half its final projected mass. All half-mass ages are given as scale factors, which means that larger values correspond to younger ages.

\section{Measuring the Assembly Bias Signal} 
\label{clustering_section}

Following most previous works, we measure the halo assembly bias signal by comparing the clustering of old versus young halos, at fixed mass. To effectively fix mass, clustering statistics must be calculated in bins of halo mass that are narrow enough so that there is no residual dependence of clustering on mass within the bin. At the same time, mass bins should be wide enough to contain a large enough number of halos for calculating robust clustering statistics. This is especially pertinent given that we will make additional cuts on halo age within the mass bins, thus further reducing the halo sample size. In this work, we adopt eight logarithmic bins of mass from $10^{10} \ h^{-1} \ \MS{}$ to $10^{11.6} \  h^{-1} \ \MS{}$ with each bin having a width of $0.2$ dex. Within each bin, we rank halos by their half-mass age and construct `old' and `young' samples that contain the 25\% oldest and 25\% youngest halos. We find that this choice of binning strikes a good balance between the competing needs to control for mass while also having large enough samples of halos for robust measurements of clustering. The age samples in our largest mass bin contain 152,966 halos.

We use all three definitions of halo mass in this analysis, described in the previous section: the actual simulation mass, the model mass that results from the fit to the entire mass accretion history, and the projected mass that is extrapolated from the fit to the early portion of the halo's history. Each definition of mass is accompanied by its own definition of half-mass age, as described in the previous section. For each definition of mass we construct one sample containing all halos in a mass bin, as well as the 25\% oldest and 25\% youngest samples. Thus for each mass bin, we have nine sets of halos for which to measure clustering.

To measure the clustering of a given halo sample, we calculate the two-point correlation function, $\xi(r)$, which is the excess probability of finding a pair of halos separated by a physical distance $r$, over what would be expected for a random distribution. To calculate $\xi(r)$, we use the equation
\begin{equation}\label{2pcf}
\xi(r) = \frac{DD(r)}{RR(r)} - 1,
\end{equation}
where $DD(r)$ is the number of halo pairs at separation distance $r$ in the data, and $RR(r)$ is the number of pairs at separation distance $r$ in a random set of data of equivalent size and volume. We use the CorrFunc package \citep{Corrfunc}, a suite of highly optimized and robustly tested codes for calculating clustering statistics. Since the Vishnu volume is a periodic cube, $RR$ is calculated analytically\footnote{For more information, see: 
\url{https://corrfunc.readthedocs.io/en/master/modules/rr_autocorrelations.html\#rr-autocorrelations}}. 

To facilitate the interpretation of our results, we compute relative bias factors for old and young halos in each mass bin. First we divide $\xi(r)$ for old halos by that for all halos and take the square root of that ratio to obtain a relative bias as a function of scale, $b(r)$. Since assembly bias is a large scale effect and bias functions become scale independent at scales larger than a few Mpc \citep{Scherrer_Weinberg_1998,Narayanan_2000}, we average the values of bias in all bins corresponding to physical separations in the range $r: 5-30h^{-1}$Mpc. This yields a single relative bias factor $b$ for the 25\% oldest halos in a mass bin. We then do the same thing for the young halos and for all our mass bins. The end result is a bias for old and young halos as a function of halo mass, and for each age definition. An assembly bias signal corresponds to a value of $b$ that is different from unity. Specifically, $b>1$ for old halos and $b<1$ for young halos \citep[e.g.,][]{Salcedo2018}.

To obtain uncertainties for our measurements, we assume that the main source of noise in the correlation function is Poisson noise from the $DD$ term (since $RR$ is calculated analytically). The Poisson uncertainty in the pair counts is $\sqrt{DD}$. We then propagate this error all the way to our final bias measurements.

\section{Results} \label{Results}

\begin{figure*}
    \centering
    \includegraphics[width=0.7\textwidth]{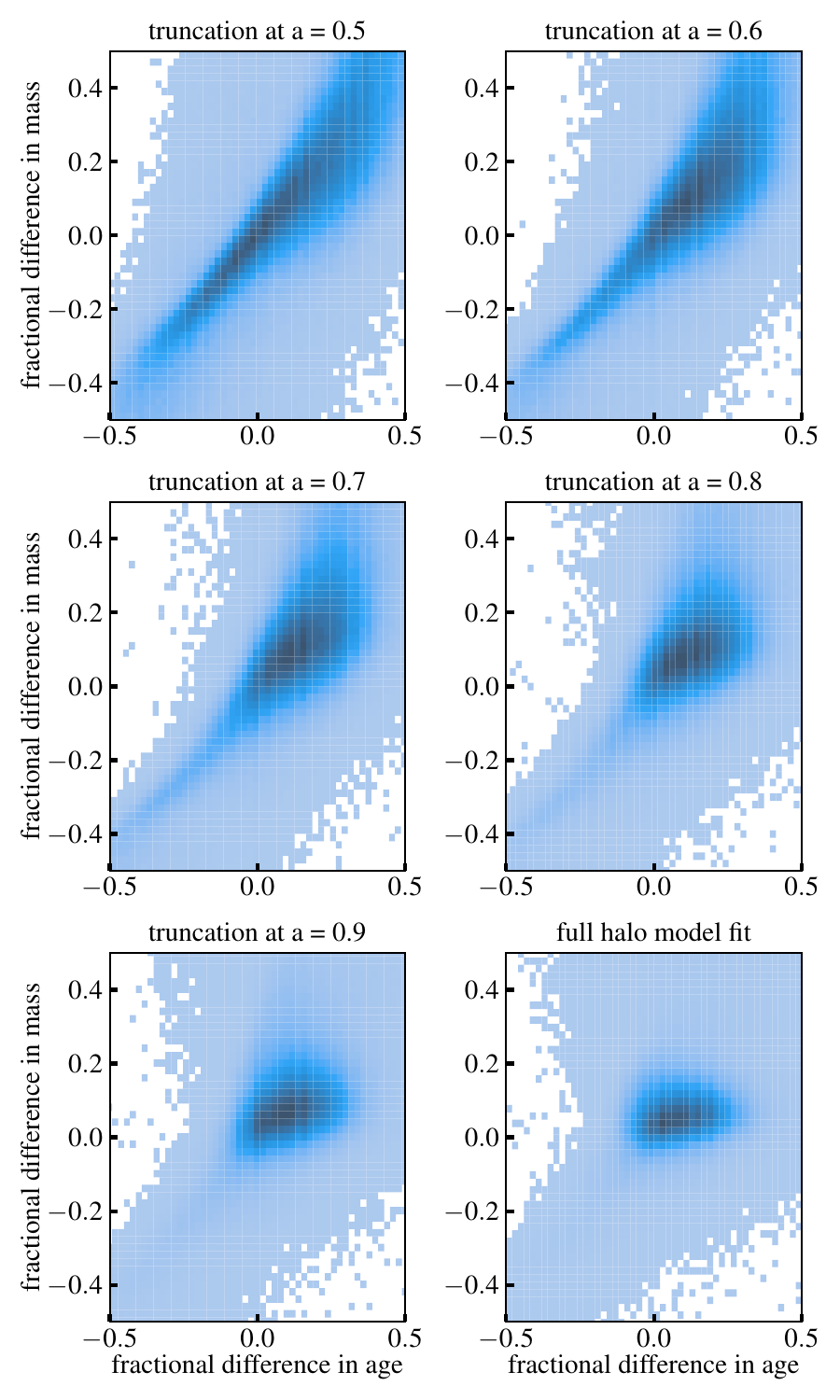}
    \caption{Bivariate distributions of fractional age and fractional mass differences between fitted and actual halos. Each panel shows fits for different projected scale factor from $a=0.5$ (top left) to $a = 1$ ("model fit", bottom right). Each panel shows the bivariate distribution of the fractional difference in the age (x-axis) vs. fractional difference in mass (y-axis). Fractional differences are defined by taking the mass (or half-mass scale factor) from the fit minus the actual mass (or half-mass scale factor) from the simulation. Arrested development halos are thus expected to appear in the top-right quadrant of each panel, since their fitted masses and half-mass scale factors should be larger than their actual values (see the example in Fig.~\ref{fig:regular_and_AD_halo_fits}). As expected, the fractional differences in mass and age decrease for most halos when fits are done adopting later truncation times.}
    \label{fig:Mass_age_comparison}
\end{figure*}
\begin{figure*}
    \centering
    \includegraphics[width=1\textwidth]{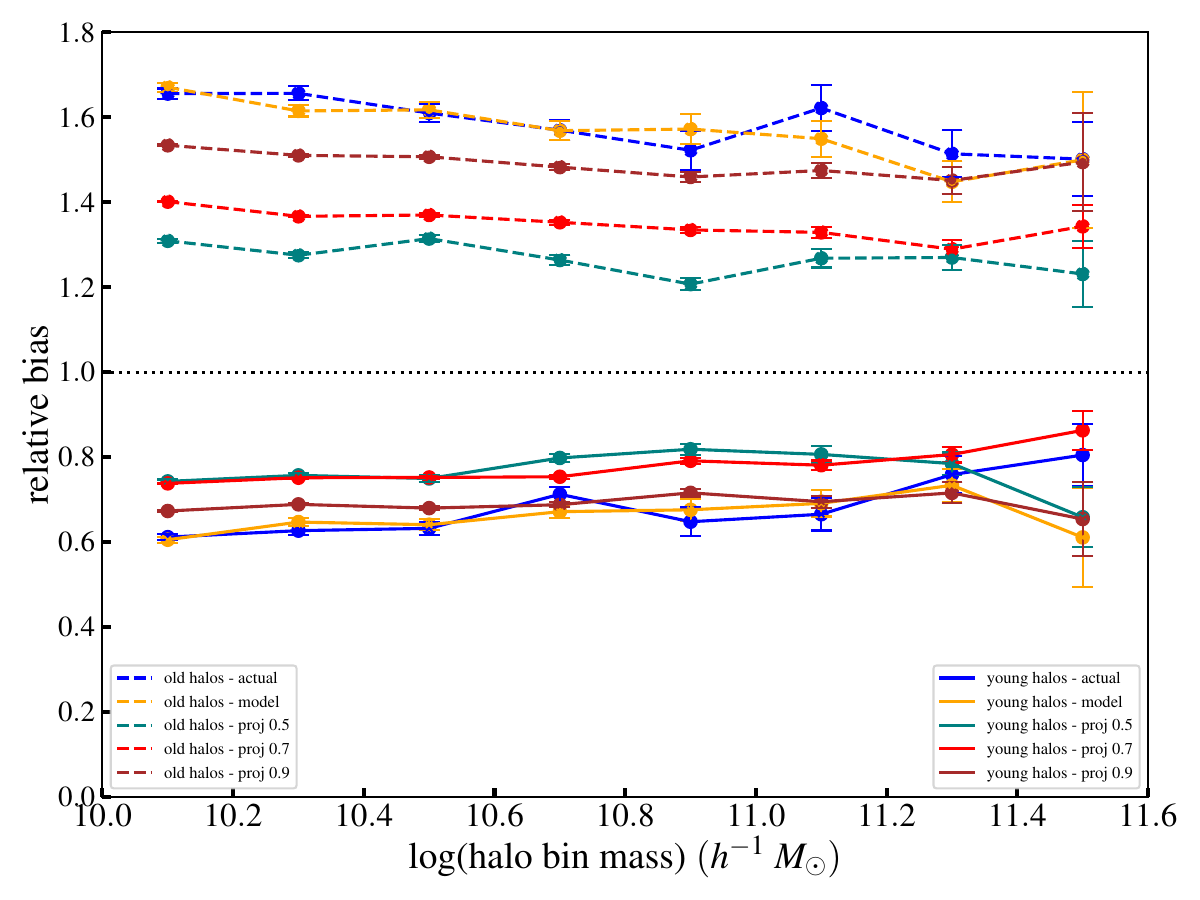}
    \caption{Relative bias of halo populations (y-axis) as a function of final halo mass (x-axis). Dashed lines show the large-scale clustering bias of the oldest 25\% halos in a set of eight mass bins divided by the bias of all halos in the same bins. Solid lines show the same, but for the 25\% youngest halos in each mass bin. The difference between the relative bias of old and young halos at fixed mass represents the halo assembly bias signal. Blue lines show results for when actual simulation halo masses and ages are used, while the other lines show results for when halo masses and ages come from fits to the halos' mass accretion histories. Yellow lines show results for fits to halos' entire histories, while brown, red, and cyan lines show results for fits out to progressively earlier truncation scales, but extrapolated to $a=1$. This represents our attempt to correct for arrested development. The cyan lines show results for the earliest truncation scale of $a=0.5$. There is a significant reduction in the assembly bias signal when halos' masses and ages are predicted from their early histories.}
    \label{fig:Bias_vs_Mass}
\end{figure*}

Figure \ref{fig:Mass_age_comparison} shows the resulting masses and ages from our fits compared to the actual masses and ages from the simulation, for all halos in our analysis. Specifically, the figure shows the bivariate distribution of the fractional difference in age (fitted half-mass scale factor minus actual half-mass scale factor) on the x-axis vs. the fractional difference in mass (fitted mass minus actual mass) on the y-axis. Each panel shows results for fits done using a different fit truncation point, from $a=0.5$ in the top-left panel, to $a=1$ in the bottom-right panel. Halos for which fits produce masses and ages that match their actual masses and ages (for example, the "normal" halo shown in the top panel of Fig.~\ref{fig:regular_and_AD_halo_fits}) appear in the middle of each panel, close to (0,0). On the other hand, arrested development halos for which fits should produce masses that are larger and ages that are younger than their actual masses and ages (like the halo shown in the bottom panel of Fig.~\ref{fig:regular_and_AD_halo_fits}), appear in the top-right quadrant of each panel.

Figure~\ref{fig:Mass_age_comparison} shows several interesting features. First we can see that the distributions of mass and age differences have a significant amount of scatter, meaning that halos exhibit a large range of behaviors when we fit their mass accretion histories. However, some clear trends can be seen. Looking at the top left panel where fits are preformed using a truncation scale of $a=0.5$, we see that most halos lie in on a sequence running from the lower-left to the upper-right. Halos on the upper part of this sequence have extrapolated fitted masses that are larger than their actual simulation masses, and ages that are younger (i.e., larger half-mass scale factors) than their actual simulation ages. These halos have likely experienced arrested development and our fits to their early histories are possibly estimating what their growth would have been without this effect. Correcting these halos is the main goal of this study. However, there is also a significant population of halos that live in the lower-left part of the sequence. These halos have extrapolated fitted masses that are smaller and ages than are older than their actual simulation masses and ages. One likely cause for this effect may be an inability of the fitting function to account for a major merger late in the halo's history. A sudden mass spike at late times would cause the extrapolated fit to underestimate the final mass and overestimate the age. The sequence that we see in the figure thus represents different types of late-time mass growth, with halos that experience late-time mergers on the bottom-left, and arrested development halos on the top-right.

We now look at how fit results change as we shift the truncation scale to later times. From the different panels in Figure~\ref{fig:Mass_age_comparison}, we see that the scatter is reduced for both mass and age, and the strong sequence we see for a truncation scale of $a=0.5$ gradually disappears. For truncation scales close to $a=1$, most halos have fitted "model" masses and ages that are close to their actual values from the simulation. This behavior is expected since the fits in these cases take the halos' late-time evolution into account. Of particular note is that even for the case of the full model fit (the bottom right panel of figure \ref{fig:Mass_age_comparison}), in which halos are fit over their entire history, most halos have fractional mass and age differences that skew slightly positive. This means that our fitting function overpredicts the final mass of halos by $\sim 5\%$, and overpredicts the half-mass scale factor of halos by $\sim 10\%$. This could be a failure of our simple fitting function to properly describe the global shape of halo mass accretion histories, or it could be a systematic effect in the fits caused by small-scale variability in the accretion histories. Either way, we need to make sure that this small systematic effect does not impact our assembly bias analysis. 

The primary result of this work is shown in Figure \ref{fig:Bias_vs_Mass}, where we investigate how the assembly bias signal changes after "correcting" halo masses and ages. The dashed and solid lines in Figure~\ref{fig:Bias_vs_Mass} show the relative halo bias as a function of halo mass for the 25\% oldest and 25\% youngest halos, respectively. The difference in clustering between old and young halos at fixed mass is the assembly bias signal. Blue lines show results using the actual halo masses and ages from the simulation. This shows what previous authors have found \citep[e.g.,][]{Salcedo2018}, though our results extend to significantly lower halo masses. We find that the oldest 25\% halos have a large-scale clustering bias that is $\sim 1.65$ times greater than that of all halos, while the 25\% youngest halos have a large-scale clustering bias that is $\sim 0.6$ times that of all halos, at the smallest masses we study. The effect decreases slightly in higher mass bins.

The other sets of lines in Figure~\ref{fig:Bias_vs_Mass} show results when we use halo masses and ages from our accretion history fits. The yellow lines show results for fits that are made over the entire accretion history (i.e., the "model" fits). We note that using these model masses and ages does not alter the assembly bias signal, as is evidenced by the fact that the yellow and blue lines are in agreement. This is important because it demonstrates that the small systematic shifting of masses and ages that we observed in Figure~\ref{fig:Mass_age_comparison} does not introduce a significant change to the assembly bias signal. Consequently, any reduction of this signal from fitting to the early part of halos' histories and extrapolating to $a=1$ can be explained by our correcting for arrested development, rather than systematic errors in the fitting process itself. The remaining sets of lines in Figure~\ref{fig:Bias_vs_Mass} show results when we use masses and ages from fits to the early parts of halo histories. The lines show results for progressively earlier fit truncation scales, with brown, red, and cyan lines corresponding to truncation scales of $a=0.9$, 0.7, and 0.5, respectively. The figure shows clearly that the assembly bias signal is substantially reduced when we predict halo growth from its early history. In the case of the earliest fit truncation scale that we consider, $a=0.5$, the signal for old halos is reduced by about a half, while the signal for young halos is reduced by a bit more than a third. The reduction in the assembly bias signal is less for fit truncation scales larger than $a=0.5$, as expected. These results are consistent with our hypothesis that arrested development is at least partially responsible for the assembly bias of low mass halos and that we can mitigate this effect by fitting the early growth of halos and extrapolating to predict what the late-time growth would have been in the absence of arrested development.

It is worth thinking about how our arrested development corrections are causing a reduction in the assembly bias signal. It is not immediately obvious how changing the masses and ages of halos will affect their clustering within age quartiles because it depends on several factors, such as (i) how the change in halos' masses and ages depends on their larger environment, (ii) how the relative change in mass and age changes a halo's location in the mass-age relation, and (ii) how the clustering of the overall halo population depends on mass.

\begin{figure}
    \centering
    \includegraphics[scale=0.5]{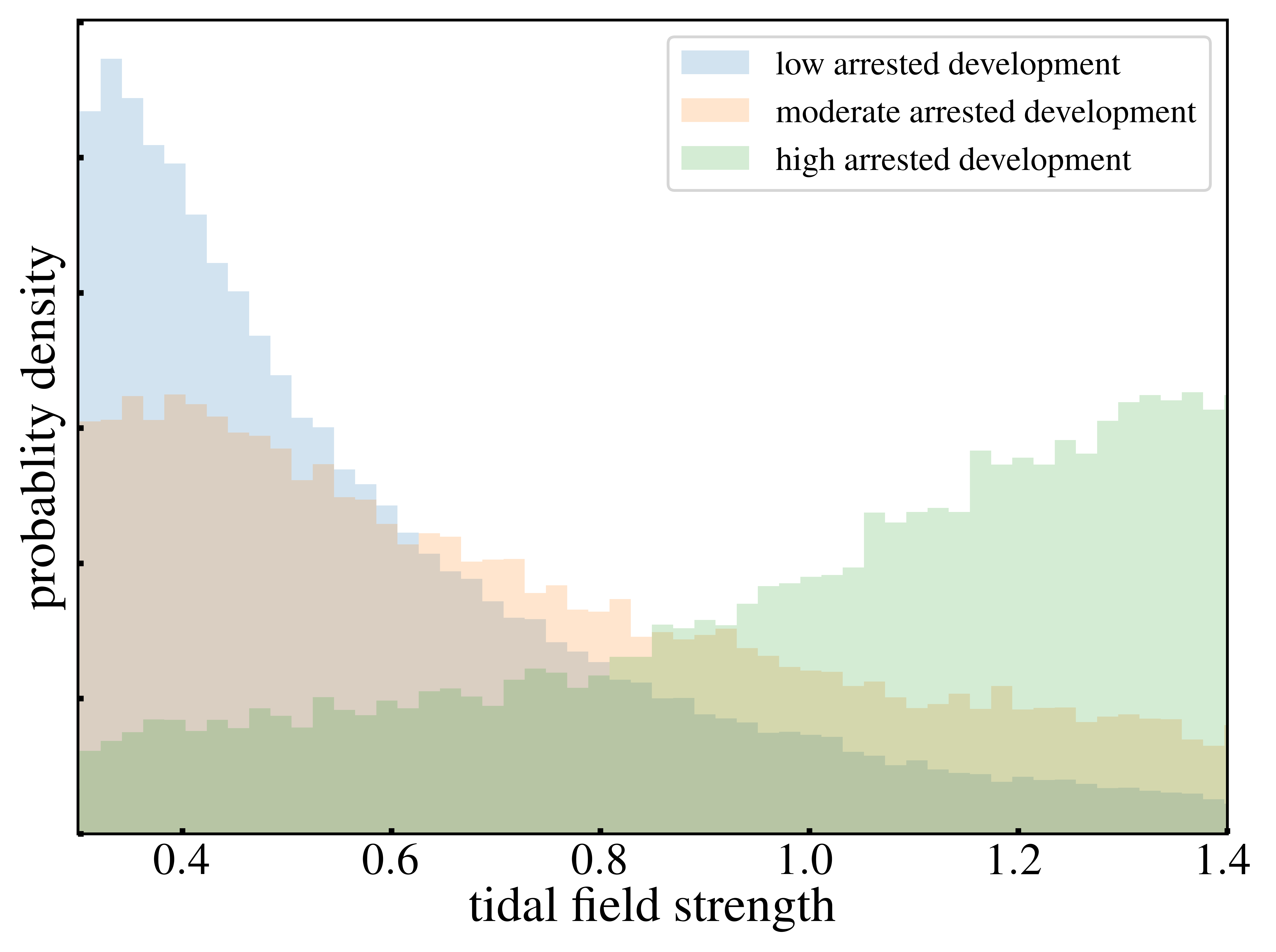}
    \caption{Distributions of tidal field strength for halos with different degrees of arrested development. The shaded histograms show the normalized distributions of tidal field strength, where the strength of the tidal field affecting each halo is defined as the ratio of the halo's radius over the Hill radius due to the nearest massive neighbor. The three histograms show distributions for halos that have experienced low, moderate, and high degrees of arrested development in their mass accretion histories. Halos that have experienced a high degree of arrested development tend to live in regions containing strong tidal fields.
    }
    \label{fig:Tidal_Field}
\end{figure}

Let us start by considering how our arrested development corrections depend on the environment. As we discussed in section~\ref{Introduction}, several previous authors have demonstrated that assembly bias for low mass halos is caused by their tidal field environment \citep[e.g.,][]{Hahn2009, Mansfield2020}. It is thus interesting to investigate the tidal fields of halos as a function of the degree of arrested development that they exhibit. To study the tidal field, we use the `tidal force' quantity provided by the ROCKSTAR halo finding algorithm. For each halo, the algorithm calculates the tidal force from all its neighbors and finds the one that exerts the maximum tidal force on the halo, which is typically its closest more massive neighbor. ROCKSTAR then outputs the value of $R_\mathrm{halo}/r_\mathrm{Hill}$, where $R_\mathrm{halo}$ is the radius of the halo and $r_\mathrm{Hill}$ is the Hill radius due to the massive neighbor, which is defined as the distance from the halo at which the gravitational influence of the neighbor exceeds that of the halo. Small values of $R_\mathrm{halo}/r_\mathrm{Hill}$ thus correspond to weak tidal fields while large values correspond to strong tidal fields. Halos that have a ratio larger than unity live in such strong tidal fields, that matter in their own outskirts is more bound to the massive neighbor than to themselves. These halos are likely in the process of being tidally stripped. To study how this tidal parameter depends on arrested development, we consider the mass accretion history fits using a projected scale factor of 0.5, and select subsamples of halos based on the fractional difference in mass and age that they have between these extrapolated fits and their original histories (the top-left panel of Fig.~\ref{fig:Mass_age_comparison}). We first make a sample where the fractional difference in mass and age is between -0.1 and 0.1. These halos exhibit little to no arrested development. We then make a "moderate" sample by selecting all halos whose fractional difference in mass and age is between 0.4 and 0.6. Finally, we make a "high" sample where the fractional difference in mass and age is greater than 1. Figure~\ref{fig:Tidal_Field} shows the normalized distributions of tidal field strength ($R_\mathrm{halo}/r_\mathrm{Hill}$) for these three samples. The figure shows a dramatic difference between the samples. Halos that exhibit a high degree of arrested development tend to live in very strong tidal field environments, while most halos with little to no arrested development live in much weaker tidal fields. Since strong tidal fields are caused by massive neighbors, it is safe to assume that halos in strong tidal fields live in highly clustered regions because they inherit the strong clustering of their massive neighbors.

\begin{figure}
    \centering
    \includegraphics[scale=0.45]{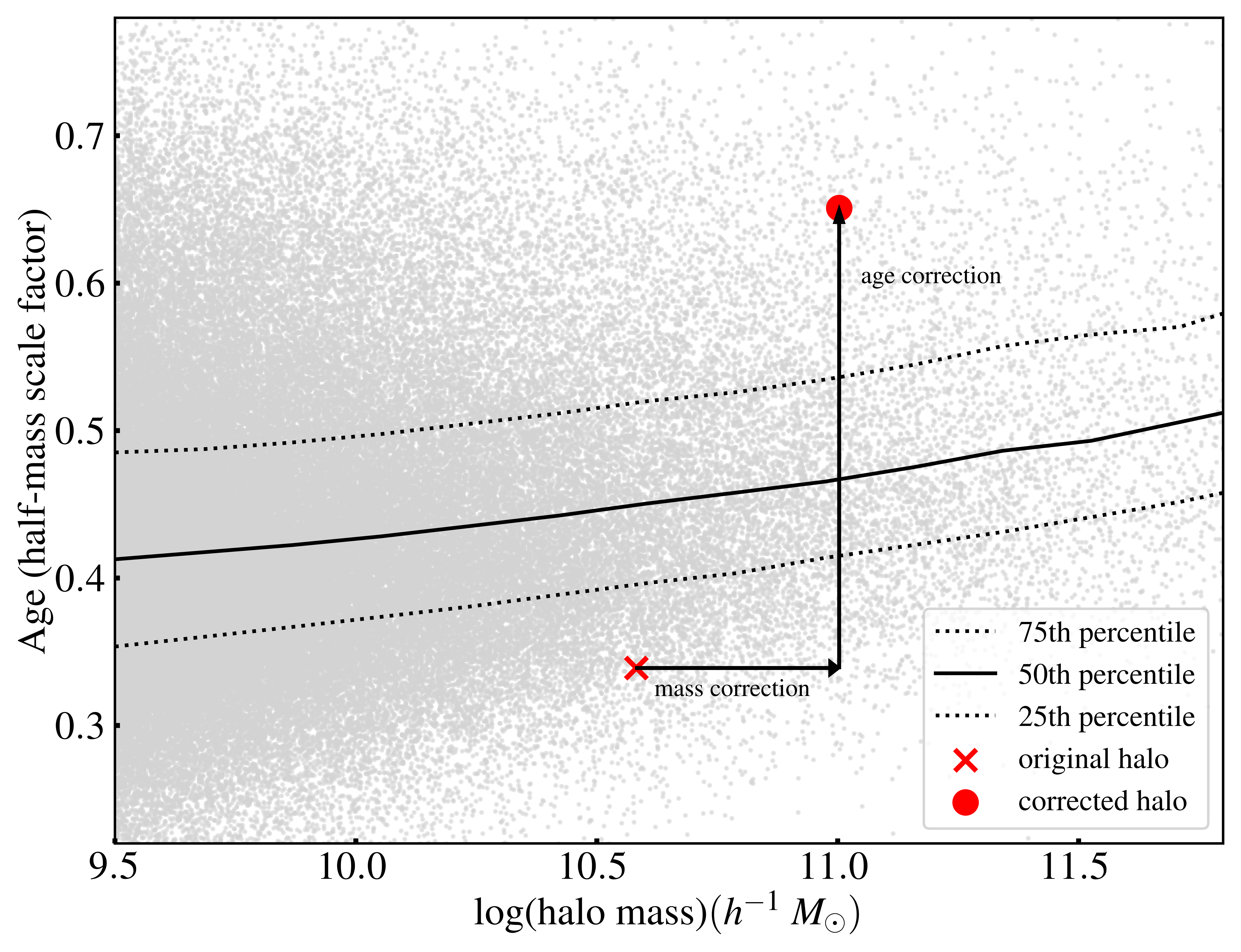}
    \caption{
    Illustration of the effect of our arrested development correction on the mass and age of a halo relative to the full halo population. Gray dots show the masses and ages (half-mass scale factors) of a random subset of all halos in our sample. The median age as a function of mass is shown by the solid black line, while the 25th and 75th percentiles are shown by the dotted lines. The red X symbol shows the original mass and age of the example halo whose mass accretion history is shown in the bottom panel of Fig.~\ref{fig:regular_and_AD_halo_fits}. The red circle shows the adjusted mass and age of this halo after applying our arrested development corrections and the arrows illustrate the mass and age corrections separately. Correcting for arrested development results in the halo having a higher mass and younger age, such that it ends up if a different age quartile for its mass.
    }
    \label{fig:Age_mass_pcts}
\end{figure}

Let us now examine how our mass and age corrections change the location of a halo in the mass-age relation. Figure~\ref{fig:Age_mass_pcts} shows the masses and ages (half-mass scale factors) of all our halos (gray points), as well as the median, 25th and 75th percentiles of age as a function of mass (solid and dotted black lines, respectively). The plot shows the well-known mass-age relation for halos whereby more massive halos tend to be younger (i.e., have assembled their mass more recently) than less massive halos. The red X symbol on the plot shows the original location of the example halo whose history is shown in the bottom panel of Figure~\ref{fig:regular_and_AD_halo_fits}. This halo's age places it in the oldest age quartile for its mass. The fact that this halo likely lives in a strong tidal field and is thus in a highly clustered region perfectly demonstrates the assembly bias effect: old halos are more clustered than young halos at fixed mass. Now we consider the mass and age corrections that we apply with the intention of "erasing" arrested development. First, the mass of this halo will increase, which is shown by the horizontal arrow. The effect of this change on the assembly bias signal is complicated. Due to the sloped mass-age relation, this halo that lives in a highly clustered region is now even older in the age ranking of halos at fixed mass. This would make the assembly bias signal even stronger than before. On the other hand, the clustering of halos increases with mass, which means that, at this new mass, the halo's clustering will not be as enhanced relative to the total population as it was before. This would make the assembly bias signal weaker than before. Since these two effects work in opposite directions, the net effect on assembly bias depends on how strongly halo clustering increases with mass. This has been studied extensively by many past works, which find that the large scale clustering of halos is fairly constant at low masses and starts to increase exponentially at masses close to the nonlinear mass $M_*$ \citep[e.g.][]{Sheth_Tormen1999, Seljak_Warren2004, Tinker2010}. The low halo masses that we are considering here are well below $M_*$, so the clustering of halos is mostly independent of mass. As a result, a mass correction that moves highly clustered halos to larger masses should have a net effect of strengthening the assembly bias signal. However, we also correct halo ages, which become correspondingly younger as we increase their masses. The vertical arrow in Figure~\ref{fig:Age_mass_pcts} shows the age correction that we apply to our example halo, which places it at a corrected location indicated by the red circle. The halo is now in the youngest age quartile at fixed mass, which will certainly contribute to a reduction of the assembly bias signal. This occurs because the ratio of the age shift to the mass shift for this halo exceeds the slope in the halo mass-age relation. The reduction in the global assembly bias signal that we see in Figure~\ref{fig:Bias_vs_Mass} takes place because the arrested development corrections that we obtain from our extrapolated mass accretion history fits behave in a qualitatively similar way as in the case of the example halo we considered here.

\section{Summary and Discussion} \label{summary}

In this paper we have tested a new method for probing halo arrested development as a cause of halo assembly bias in the low mass regime. Furthermore, our method provides a potential way to mitigate the effects of assembly bias in N-body simulations. We define halo arrested development as the late-time slow-down of mass accretion in some low mass halos due to the tidal field of the surrounding mass distribution. Arrested development thus results in halos having lower final masses and older measured ages than they would have had in the absence of arrested development. These halos live in dense large-scale environments and are thus more clustered and older than other halos of the same final mass, leading to an assembly bias signal. The new idea that we introduce and test is that we can use the early part of a halo's mass accretion history, before arrested development has kicked in, to predict what the late-time growth of the halo would have been in the absence of arrested development. We do this by fitting an analytic function to the mass accretion history at early times, for example before $a=0.5$, and extrapolating the fit to $a=1$. We then calculate a "corrected" mass and age for the halo based on this fit. Using these corrected masses and ages of halos, we check to what extent the assembly bias signal is affected. Our results show that this method leads to a substantial decrease in the assembly bias signal. This supports the hypothesis that arrested development is a major cause of assembly bias, in agreement with \citep{Mansfield2020}, and validates our new approach to mitigating the effect.

The mass accretion fitting method cuts the assembly bias signal by a half for the 25\% oldest halos and by a little more than a third for the 25\% youngest halos, in our lowest mass bins of $\sim 10^{10} h^{-1} \MS$. Though this result is promising, our method does not yet produce the magnitude of reduction that would lend it useful as a method of fully mitigating assembly bias in cosmological N-body simulations. It is thus worth thinking about why the method does not fully remove assembly bias. One possibility is that arrested development is not solely responsible for low mass assembly bias. This is certainly possible, but before reaching this conclusion, we must first gain confidence that our methodology is able to fully remove the effects of arrested development and does not introduce any additional spurious assembly bias signal. 

There are several potential reasons why our fitting method may not be fully accounting for and removing arrested development. One possibility is that the effects of arrested development begin to impact a halo at very early times, before the scale $a=0.5$ that represents the earliest fit truncation scale we adopt. Figure~\ref{fig:Bias_vs_Mass} shows a decrease in the assembly bias signal for old halos when we shift the truncation scale from $a=0.7$ to $a=0.5$ so perhaps the signal would further reduce by restricting the fit to even earlier times. However, fits might start to become unreliable if we restrict them to too small a portion of the history. Also, the assembly bias signal for young halos does not change from $a=0.7$ to $a=0.5$, making it unlikely that pushing to earlier times would affect young halos. Another possibility is that the early mass growth of a halo has limited power to predict its late-time growth. The extrapolation of our fits to $a=0$ may thus contain a fair amount of stochasticity or even systematic effects that fail to properly remove the effects of arrested development in many halos. Alternatively, it may be the case that the early growth of halos does indeed have predictive power, but that our choice of fitting function is the limiting factor. A more sophisticated model that takes into account a halo's entire merger tree, rather than just the one dimensional mass history following the most massive progenitor, might be more successful in predicting late-time growth.

A different possibility is that our fitting methodology is introducing a spurious assembly bias signal. Figure~\ref{fig:Mass_age_comparison} revealed that many halos have "corrected" masses that are smaller than the original masses from the simulation. These are likely halos that experience fast growth at late times, e.g., through a late major merger, which is the opposite of arrested development. Applying our mass and age correction to these systems might be inducing a fake assembly bias signal. An investigation of ways to identify which halos require a correction deserves further study. 
 
These are difficult issues to disentangle because arrested development is not a simple binary process, where a halo either experiences it or not. In that case, one could identify a set of "normal" halos that do not experience arrested development and use it to validate that a given correction method does not alter those halos. However, the reality is likely more complex, with every halo being exposed to some degree of tidal field and having its mass accretion impacted by its larger scale environment to some extent. One could perhaps approximately identify a sample of "normal" halos by selecting halos that do not have a nearby massive neighbor. This is worth future study. For halos that do experience arrested development, we have no way of knowing how they would have evolved in the absence of this effect, making it impossible to test how well our correction method works on a halo-by-halo basis. We can only test this in the aggregate by examining the assembly bias signal of all halos -- if the assembly bias signal vanishes then our correction method must be working. Of course, this only holds true under the assumption that arrested development is the sole cause of assembly bias. The main goal of this work was to test a new idea for probing and mitigating the effect of assembly bias. Our preliminary results are encouraging and thus warrant further study along the various lines discussed in this section.

\section*{Acknowledgements}
Parts of this research were conducted at Vanderbilt University through the Fisk-Vanderbilt Bridge to Ph.D. Program and the Establishing Multimessenger astronomy Inclusive Training (EMIT) program. Parts of this research were conducted by the Australian Research Council Centre of Excellence for All Sky Astrophysics in 3 Dimensions (ASTRO 3D), through project number CE170100013. Computational facilities used in this project were provided by the Vanderbilt Advanced Computing Center for Research and Education (ACCRE). This research made use of NumPy \citep{harris2020array}, SciPy \citep{SciPy}, Matplotlib, a Python library for publication quality graphics \citep{matplotlib_hunter}, and corrfunc \citep{Corrfunc}. We thank the anonymous referee for asking probing questions that led to an improvement of the paper.

\section*{Data Availability}

The data underlying this article is available on request and will be shared on reasonable request to the corresponding author.



\bibliographystyle{mnras}
\bibliography{MAF_AB}


\bsp	
\label{lastpage}
\end{document}